# SECURITY PROTOCOL REVIEW METHOD ANALYZER (SPRMAN)

A.S.Syed Navaz
Department of Computer Applications,
Muthayammal College of Arts & Science, Namakkal. India

H.Iyyappa Narayanan,
Department of Computer Applications,
Muthayammal College of Arts & Science, Namakkal. India

R.Vinoth
Department of Computer Applications,
Muthayammal College of Arts & Science, Namakkal. India

*Abstract*— **This Paper is designed using J2EE (JSP, SERVLET), HTML as front end and a Oracle 9i is back end. SPRMAN is been developed for the client British Telecom (BT) UK., Telecom company. Actually the requirement of BT is, they are providing Network Security Related Products to their IT customers like Virtusa, Wipro, HCL etc., This product is framed out by set of protocols and these protocols are been associated with set of components. By grouping all these protocols and components together, product is been developed.**
**After framing out the product, it is been subscribed to their individual customers. Once a customer subscribed the product, then he will be raising a request to the client (BT) for updating any policy or component in the product. The customer has been given read/write access to the subscribed product. The customer user having read/write access is only allowed to raise a request for the product, but not the user having only the read access. The group of request is been managed as manage work queue in client area. Management of this protocol inside the product is considering as Security Protocol Review Method Analyzer. SPRMAN helps BT to overcome all the hurdles faced by them while processing the requests of their various clients using their already existing software applications. SPRMAN emphasizes on nature of the request and gives priority to issues based on their degree of future consequences. Thus SPRMAN builds a good relationship between BT and its customers.**

Introduction

The purpose of this software specification is to establish the major requirements and specification necessary to develop the software systems for the developers. The goal of this Sprman helps bt to overcome all the hurdles faced by them while processing the requests of their various clients using their already existing software applications. sprman emphasizes on nature of the request and gives priority to issues based on their degree of future consequences. This Paper secures & protects mobile platforms, remote users from abuse and attack and secures standalone desktop. This paper implements personal firewall, integration of local anti-virus tools, programs control alerting and reporting by providing baseline security posture for all corporate PC's.

I. EXISTING SYSTEM

Managed Enterprise Desktop Protection (**MEDP**) product is one of the security service offerings of MobileXpress product. MEDP is designed for users within the customer infrastructure, LAN users, mobile users and VPN connected users. To summarize, MEDP offers the following functionality:

- Secures and protects mobile platforms and remote users from abuse and attack
- Secures standalone desktop
- Implements personal firewall, integration local anti-virus tools, program control and alerting and reporting
- Provides baseline security posture for all corporate PCs

In order to make MEDP provide the security features, certain rules and conventions are to be implemented on the Protocol Server that will control the Desktop and ensure availability of the MEDP security features.

A. *Limitations of Existing System*

o At present, there is no system in BT that can support managing the rules and conventions for desktop protection.
o BT cannot support the requirements for MEDP.

II. PROPOSED SYSTEM:

At present, there is no system in BT that can support managing the rules and conventions for desktop protection. Though BT Infonet has a system Online Site Survey (OSS) that implements the rules and conventions for similar products like MEDP, but it cannot support the requirements for **MEDP**. Therefore it was decided to develop a new system named **SPRMAN** for analyzing & managing rules and conventions for MEDP and other related security products.

Security Protocol Review Method Analyzer, a telecom domain Paper, is being developed for the client British Telecom (**BT**) UK Telecom Company. Actually the requirement of BT is, they are subscribing a network security related products to their IT customers like Virtusa, Wipro, HCL etc., This product is framed out by set of protocols and these protocols are been associated with set of components. By grouping all these protocols and components together, product is been developed.

  A. *Advantages of Proposed System*
- After framing out the product, it is been subscribed to their individual customers.
- Once a customer subscribed the product, then he will be raising a request to the client (BT) for updating any component in the product.
- SPRMAN ensures that requests have been submitted securely i.e. changes are transferred via secure mechanism to BT.
- Only valid members in the organization can submit changes.
- It enables only specified sites to customers of BT and provides baseline security posture for all corporate PCs.

III. PROBLEM DESCRIPTION

Managed Enterprise Desktop Protection (**MEDP**) product is one of the security service offerings of Mobile-Xpress product. MEDP is designed for users within the customer infrastructure, LAN users, mobile users and VPN connected users. To summarize, MEDP offers the following functionality: Secures and protects mobile platforms; remote users from abuse; attack and secures standalone desktop; Implements personal firewall; integration local anti-virus tools; program control and alerting and reporting; Provides baseline security posture for all corporate PCs.

In order to make MEDP provide the security features, certain rules and conventions are to be implemented on the Protocol Server that will control the Desktop and ensure availability of the MEDP security features. At present, there is no system in BT that can support managing the rules and conventions for desktop protection. BT cannot support the requirements for MEDP. SPRMAN helps BT to overcome all the hurdles faced by them while processing the requests of their various clients using their already existing software applications. SPRMAN emphasizes on nature of the request and gives priority to issues based on their degree of future consequences. Thus SPRMAN builds a good relationship between BT and its customers.

After framing out the product, it is been subscribed to their individual customers. Once a customer subscribed the product, then he will be raising a request to the client (BT) for updating any policy or component in the product. The customer has been given read/write access to the subscribed product. The customer user having read/write access is only allowed to raise a request for the product, but not the user having only the read access. The group of request is been managed as manage work queue in client area. Management of this protocol inside the product is considering as Security Protocol Review Method Analyzer.

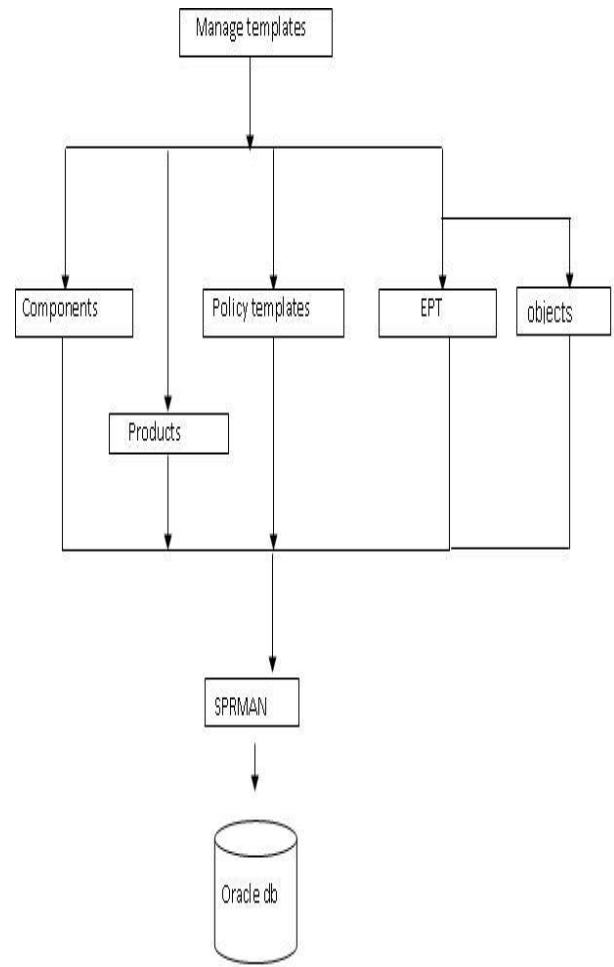

Figure 2: Security Protocol Review Method Analyzer – Module operation

IV. MODULE DETAILS

*MANAGE TEMPLATES:*

This user case describes how the Ops users manage the product and protocol templates.

A. *Pre – Conditions:*

User logged in to the system. User is an Ops User and has privilege to work on any Product Template. This module is only for Ops users.

B. *Business Rules:*

A product has policies. A product has basic features as defined in the Product components. A product can have features specific to a customer. A Protocol is specific to a product. A protocol can have many protocol objects. A set of protocol objects form a protocol package. PEP has features that define basic properties and associated Package(s) for the PEP.

C. *Main Event List / Flow of Events:*

A product can be defined any time within the system. Associate the product with a one or more number of policies. A product's components can be defined from the Product Protocol Generic Items. A Product has one or many Protocol Enforcement Points. Each PEP has a PEP Template. A protocol template is a form for capturing the data heads for capturing protocol data items. A protocol template format is used for capturing protocol change request data for protocol objects.

Once a Protocol Template is created, a corresponding protocol object table will be created by using the protocol template components. A protocol template cannot be deleted if any one customer is still using this template. The Ops user should be able to define Parent-Child relationship between the items in the template so that the child item is displayed in lower level than the Parent item. The child item may inherit the attributes of the parent item. Changes to the child item can only be made when the Parent is enabled for changes. It should be possible to define an item with File data type so that upload/download of the file is possible with the template.

*CUSTOMER MASTER:*

This user case describes how the Ops users manage the workspace for a Customer.

a. *Pre-conditions:*

User logged in to the system. User is an Ops User and has privilege to work on any customer request.

b. *Business Rules:*

An Ops User can create a new customer account in absence of an interface with Wizard. A protocol enforcement point (PEP) is associated with each Product. Each PEP Id is unique for the customer in the system.

When a customer gets the product, it gets all policies associated with the Product. Customer's user will be the recipient of all email communications for the Customer. A customer or product record cannot be deleted, if has got child records.

c. *Main Event List / Flow of Events:*

The Ops user can create a customer record. The information required is in two sets, one specific to the customer and one set per customer product.

A minimum of one PEP Id is to be created per Customer. A PEP Template will be used to capture features of PEP. Associate a Product with the Customer. Create Customer Policies by associating all policies available for the Product with the Customer. Create default protocol objects (blank objects) for all Customer Policies. Associate all these policies into one Protocol Package. Create a PEP Request for the customer. Create a Protocol Package request for the customer. For existing customers, Ops user can add new PEP Ids upon customer request. All new PEPs will be associated with either existing protocol package or a new Package (Null Protocol objects).

If new, Ops user will define the package upon approval of such request. The package will have null protocol objects. All new PEPs will have a user defined PEP feature. A Change Request will be created for adding new PEP for a Customer. One Request will be created per Protocol Package.

*USER MASTER:*

This user case describes how the any user record can be defined either by the Ops User.

a. *Pre-conditions:*

Admin User logged in to the system. User has the write access required to create a user in the system.

b. *Business Rules:*

Ops user can create a user account. A user can be given a role as Customer user or Ops user. A Customer user can be associated with the product with Read-Write or Read-Only privilege. An Ops user is given Read-Write access to all the products/customers. A Customer user must have the following item in the profile: Email (which is the same as user account) and Off-Line Verification information. The Off-Line Verification is used for Ops user to verify the Customer user when they call. This information is private to other customers but visible to the Ops users. On first login, the Customer user must set Off-Line Verification information. The Off-Line Verification questions are "Year/Month when you entered the company" and/or "Your home Zip code". A Customer user must be able to enter or edit the profile including following items: Name, Phone number.

   c. *Main Event List / Flow of Events:*

Define a new user. Associate the user with Customer and Product. A user having the RW privilege can create a change request for any protocol within any package for its product(s). The list of Product that the user is associated should be shown the user upon login so that they can choose with Product they want to work on

*WORK QUEUE:*

This user case describes how the Ops users manage the work queue created when a new customer signs in or an existing customer requests change(s) to its existing policies.

   a. *Pre-conditions:*

User logged in to the system. User is a Ops User and has privilege to work on any customer request. This may be the landing page when the Ops user logs into the system.

   b. *Business Rules:*

An admin op user can manage (assign, reassign, reject, suspend) any new request. A non admin ops user can assign any new request to him/her only. A non admin user can generate report on his assignments only.

An admin ops user can generate report on any ops user. Valid values for Status are: Saved, Submitted, Cancelled, Under Review, Rejected, Pending, Approved, and Completed.

   c. *Main Event List / Flow of Events:*

When Ops user logs into the system, the summary of requests will be displayed. The items on the screen are: Request Id, Class Of Service, Request Time, Customer ID-Name, Product, Status, Assigned to. The user can sort on any key. The filter is available on Status field. The default sort order is by descending Request Id.

The user can opt to see certain type of records (New, Assigned, Suspended etc). The user can click on the customer to see all details of the work packages under a request. This will show PEP Ids, Policies, Status, Requested Date and time, If a particular Protocol is clicked, the Protocol Object details of that Protocol should be displayed. Each request id has a start date and an end date.

V. SYSTEM DESIGN

A system is a combination of resources working together to convert input to usable output. Design is a process of translating requirements designed during analysis into several different alternatives for user consideration. In system design a variety of tools such as DFD, Data Dictionary-R diagram etc is used.

System Design is a process of creating alternative solution, which satisfied the study goals, evaluating and then drawing up specification of the chosen alternatives in the design phase. The detailed design of the system selected in the study is accomplished and used specification is converted into technical specification.

A. *INPUT DESIGN*

It is the process of converting the user-originated input to computer-based format. In the system design phase the expanded dataflow diagram, identifies logical dataflow, data stores and destination. Input data is collected and organized into groups of similar data.

The goal of designing input data is to make the data entry easy and make it free from logical errors. The input entry to all type of clients are user name, password only if they are valid, the client is allowed to use the software.

Thus a careful design of input stages has been taken place by giving attention to error handling, controls, and batching and validation procedures.

B. *OUTPUT DESIGN*

Computer output is the most important and direct source of information to the user. Efficient intelligible output design should improve the system's relationship with the user and help in decision-making.

Output design generally refers to the results entered by the system. For many end users on the basis of them. Output they evaluate the usefulness of the application.

VI. SYSTEM IMPLEMENTATION

Implementation is the stage of the Paper when the theoretical design is turned out into a working system. Thus it can be considered to be the most critical stage in achieving a successful new system and in giving the user, confidence that the new system will work and be effective. The implementation stage involves careful planning, investigation of the existing system and it's constraints on implementation, designing of methods to achieve changeover and evaluation of changeover methods.

Implementation is the process of covering a new system design into operation. It is the phase that focuses on user training, site preparation and file conversion for installing a candidate system. The important factor that should be considered here is that the conversion should not disrupt the functioning of the organization.

## VII. DATABASE DESIGN

### A. MANAGE TEMPLATES

| Field Name | Field Type | Possible Values | Default Value | Validation | Data source |
|---|---|---|---|---|---|
| Component Name | Label | N/A | N/A | Non editable | ComponentVO.componentName |
| Description | Table Text | N/A | N/A | The Description field should not be empty | ComponentVO.componentDesc |
| Data Type | Combo, | Text, Numeric, File | Text | N/A | Hardcoded in JSP |
| Component Type | List | Product and Policy | N/A | N/A | ComponentVO.componentFlag |
| Data Source | Combo | None, Policy Object | N/A | N/A | ComponentVO.dataObj |

| User Action | Algorithm | Next Screen Success | Parameter |
|---|---|---|---|
| Click on any column header | onClick="sortComponentDetails()" | Manage Components.Jsp | Selected Column name |

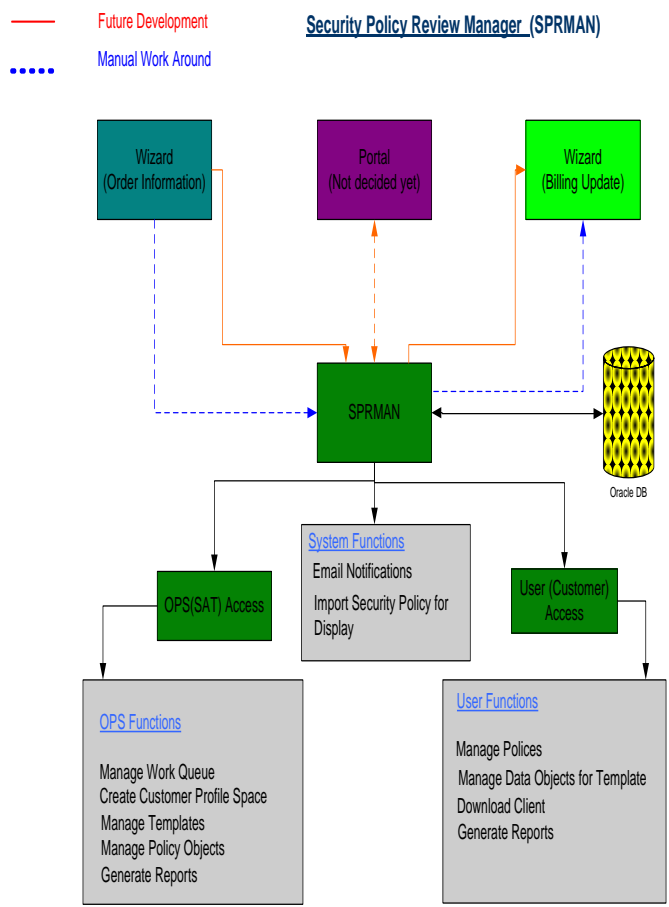

Figure 2: Security Protocol Review Method Analyzer – Design Overview

### B. PAGE NAVIGATION

## VIII. CONCLUSION

It is concluded that the application works well and satisfy the needs. The application is tested very well and errors are properly debugged. The application works well both in single as well as multiple clients. This product is framed out by set of protocols and these protocols are been associated with set of components. By grouping all these protocols and components together, product is been developed. The various features in proposed system are:

- After framing out the product, it is been subscribed to their individual customers.

- Once a customer subscribed the product, then he will be raising a request to the client (BT) for updating any component in the product.

- SPRMAN ensures that requests have been submitted securely i.e. changes are transferred via secure mechanism to BT.

- Only valid members in the organization can submit changes.

- It enables only specified sites to customers of BT and provides baseline security posture for all corporate PCs.

### REFERENCES

1. Tracie Karsjens, Stefan Zeiger, "Java Server Programming".
2. Written by the team at Visualbuilder.com, "Creating customer JSP tag Library".
3. James Goodwill, "Developing Java Servers".
4. kevin loney, george coach, "Oracle9i: The Complete Reference".

## AUTHOR'S INFORMATION

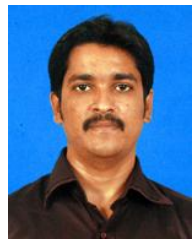 **A.S.Syed Navaz** received BBA from Annamalai University, Chidambaram 2006, M.Sc Information Technology from KSR College of Technology, Anna University Coimbatore 2009, M.Phil in Computer Science from Prist University, Thanjavur 2010, M.C.A from Periyar University, Salem 2010 and Pursuing his Ph.D in the area of Wireless Sensor Networks. Currently he is working as an Asst.Professor in the Department of Computer Applications, Muthayammal College of Arts & Science, Namakkal. His area of interests are Wireless Communications, Computer Networks and Mobile Communications.

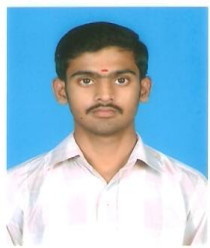 H.Iyyappa Narayanan, pursuing his BCA in Muthayammal College of Arts & Science, Namakkal. His area of interest are Computer Hardware.

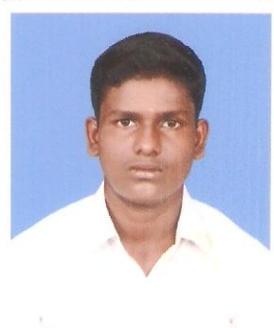 R.Vinoth,, pursuing his BCA in Muthayammal College of Arts & Science, Namakkal. His area of interest are Computer Hardware.